\documentclass[%
aip,
amsmath,amssymb,
reprint,%
]{revtex4-2}

\usepackage{graphicx}
\usepackage{dcolumn}
\usepackage{bm}

\usepackage[utf8]{inputenc}
\usepackage[T1]{fontenc}
\usepackage{mathptmx}
\usepackage{etoolbox}
\usepackage{xcolor}

\makeatletter
\def\@email#1#2{%
	\endgroup
	\patchcmd{\titleblock@produce}
	{\frontmatter@RRAPformat}
	{\frontmatter@RRAPformat{\produce@RRAP{*#1\href{mailto:#2}{#2}}}\frontmatter@RRAPformat}
	{}{}
}%
\makeatother
\begin{document}
	
	\preprint{AIP/123-QED}
	
	\title[Title]{Shear induced tuning and memory effects  in colloidal gels of rods and spheres}
	\author{Mohan Das}
	\altaffiliation[Now at ]{Department of Chemical and Biomolecular Engineering, Case Western Reserve University, 10600 Euclid Ave., Cleveland, USA, 44106.}
	\affiliation{ 
		IESL-FORTH,GR – 71110, Heraklion, Greece
	}%
	\affiliation{%
		Department of Material Science and Technology, University of Crete, GR – 71110, Heraklion, Greece.
	}%
	\author{George Petekidis}%
	\email{georgp@iesl.forth.gr}
	\affiliation{ 
		IESL-FORTH,GR – 71110, Heraklion, Greece
	}%
	\affiliation{%
	Department of Material Science and Technology, University of Crete, GR – 71110, Heraklion, Greece.
	}%
	
	
	\date{\today}
	
	\begin{abstract}
		Shear history plays an important role in determining the linear and nonlinear rheological response of colloidal gels and can be used for tuning their structure and flow properties. Increasing colloidal particle aspect ratio lowers the critical volume fraction for gelation due to an increase of the particle excluded volume. Using a combination of rheology and confocal microscopy we investigate the effect of steady and oscillatory pre-shear history on the structure and rheology of colloidal gels formed by silica spheres and rods of length \textit{L} and diameter \textit{D} (\textit{L/D} = 10) dispersed in 11 M CsCl solution. We use a non-dimensional Mason number, Mn (= F$_{visc.}$/F$_{attr.}$) to compare the effect of steady and oscillatory pre-shear on gel viscoelasticity. We show that after pre-shearing at intermediate Mn, attractive sphere gel exhibits strengthening whereas attractive rod gel exhibits weakening. Rheo-imaging of gels of attractive rods shows that at intermediate Mn, oscillatory pre-shear induces large compact rod clusters in the gel microstructure, compared to steady pre-shear. Our study highlights the impact of particle shape on gel structuring under flow and viscoelasticity after shear cessation. 
	\end{abstract}
	
	\maketitle
	
		%
	
	\section{\label{sec:level1}Introduction}
	
	Colloidal gels are soft systems formed by three-dimensional percolated network of particle clusters that are in turn formed as a result of sufficiently strong interparticle attraction (U $>>$ kT).\cite{mewis2012colloidal,petekidis2021rheology} Such a particle network, even at very low volume fractions can exhibit solid-like behaviour at rest and yield (change from solid to liquid) at large imposed stress or strain.\cite{rueb1997viscoelastic,barnes1999yield,bonn2017yield} Additionally, these systems exhibit flow behaviours such as shear thinning and thixotropy.\cite{buscall1987rheology,mewis2012colloidal,mewis2009thixotropy,larson2019review} This property of colloidal gels is extremely useful in engineering the flow properties of material systems employed in various domestic, industrial, construction and biological applications.\cite{gallegos1999rheology,gibaud2012new,mezzenga2013self,storm2005nonlinear,tan2018direct,lootens2004gelation,mewis2021rheology} Depending on the type of flow, waiting time, strength and range of attraction and solvent properties, colloidal gels may exhibit a variety of rheological behaviours. It is well established that shear flow in colloidal gels can induce structural changes such as cluster break-up,\cite{mewis2012colloidal,koumakis2015tuning,moghimi2017colloidal} shear-induced aggregation,\cite{zaccone2011shear} although phenomena such as wall-slip may also be present obscuring interpretation.\cite{walls2003yield,ballesta2013slip,laurati2011nonlinear} In addition, such structural transitions can lead to changes in viscosity and elasticity, delayed\cite{sprakel2011stress,lindstrom2012structures,landrum2016delayed} and two-step yielding\cite{laurati2011nonlinear,koumakis2011two,chan2012two} and structural collapse under gravity.\cite{manley2005gravitational} Colloidal gels are also sensitive to confinement effects leading to the formation of large space-filling particle clusters aligned in the flow-vorticity plane.\cite{osuji2008highly,grenard2011shear,varga2018large,varga2019hydrodynamics,das2020shear} As shear history affects the microstructure, the related rheological properties, such as the linear viscolelasticity, can either exhibit temporary hysteresis, termed thixotropy,\cite{petekidis2021rheology,buscall1987rheology,mewis2012colloidal,larson2019review,divoux2013rheological,jamali2019multiscale,jamali2019materials} or in some cases permanent (or extremely long-lived) changes in the microstructure and therefore resulting in a  true tunability of the viscoelastic properties.\cite{koumakis2015tuning,mewis2021rheology} The latter is expected to take place in complex yield stress systems where different  pre-shear protocols can lead to different long-lived metastable states with distinct microstructures and therefore mechanical or other properties. More recently such permanent shear induced tuning of complex yield stress fluids have been described as ``memory'' induced phenomena, as the system is considered to have incorporated a memory of its previous pre-shear history.\cite{schwen2020embedding}
	\par While at rest, depending on particle volume fraction and magnitude and range of interparticle attraction, colloidal gels can form a wide-range of structures such as fractal-like clusters, networks of particle clusters and attractive glasses.\cite{shih1990scaling,lu2008gelation,pham2002multiple,petekidis2021rheology} However, while undergoing shear flow, colloidal gel microstructure is subject to complex flow fields with competing hydrodynamic and attraction forces. Structural evolution in colloidal gels undergoing shear flow has been captured using different experimental techniques such as scattering\cite{varadan2001shear,vermant2005flow,mohraz2005orientation, masschaele2011flow,kim2014microstructure} and microscopy.\cite{hsiao2012role,masschaele2011flow,koumakis2015tuning} Moreover, significant contributions towards understanding the mechanism at particle length scale have been made using simulations.\cite{potanin1993computer,dickinson2013structure,park2013structural,colombo2014stress,park2015structural,moghimi2017colloidal} The majority of such studies are based on gels formed by particles of simple geometry such as spheres. However, most of the gel-forming materials in the natural world are formed by anisotropic particles and are of considerable interest.\cite{pignon1996structure,lieleg2009structural,capadona2008stimuli,fall2013physical,mewis2021rheology}
	\par Rod-shaped colloids (of length L and diameter D) exhibit rich phase behaviour due to their ability to form ordered states both in terms of their position as well as orientation with increasing number density.\cite{onsager1949effects,tang1995isotropic,lettinga2005kinetic,kuijk2012phase} In the dilute ($\phi\le$ (D/L)$^{2}$) and semi-dilute regime ((D/L)$^{2}$ $\le \phi \le$ (D/L), rods are not expected to exhibit any ordering and are isotropic.\cite{doi1988theory} Introducing attractions between rods in this regime leads to the formation of an arrested isotropic gel.\cite{van1998long,mohraz2004effect,solomon2010microstructural} Rods when subjected to shear flow  can flow align at shear rates defined by Pe$_r$(= $\dot \gamma$/D$_r$) $>$ 1 where D$_r$ ( = 3kT/$\pi \eta$L$^3$) is the rotational diffusion coefficient.\cite{broersma1960rotational} Colloidal gels of attractive rods subjected to shear flow at different Pe$_r$ may be prone to structural transitions that exhibit thixotropy, aging and yielding behaviour different compared to gels of attractive spheres. Studies focused on these parameters are limited with few experiments and simulations revealing the structure-rheology relationship.\cite{shafiei2012rheology,das2020shear} Experimental studies comparing fractal gels of spheres and rods revealed differences in transient anisotropy in gel microstructure during start-up of shear measurements.\cite{mohraz2005orientation} Here the gels of spheres show maximum structural anisotropy at the peak of stress whereas this effect was undetectable in rods. Simulation studies on attractive rods show a widening of the isotropic-nematic coexistence phase with increasing attraction strength. When subjected to shear flow, they exhibit tumbling of nematic domains with the time associated with tumbling decreasing with increasing attraction strength.\cite{ripoll2008attractive} Another simulation study reported the effect of shear-induced flow on attractive rods where flow aligned rod bundles formed when the rods were free to rotate about their contact points.\cite{stimatze2016torsional} Moreover, low volume fraction rod gels while undergoing shear flow under confinement, form large anisotropic rod clusters at very low values of Mn.\cite{lin2004elastic,das2020shear} Examining model rod-like systems with attractive interactions with a focus on microstructural evolution under flow would enhance our understanding of tunability of colloidal gels with shape anisotropy.
	\par In this paper, we examine the role of externally imposed shear rate (steady and oscillatory) in altering the linear viscoelastic response and yielding behaviour of colloidal gels made-up of attractive particles of different shape anisotropy (spheres and rods). Silica spheres and rods suspended in 11 M CsCl solution with interparticle van der Waals attraction are used as model colloidal gels. Our studies reveal that there is a marked difference in the response to pre-shear of attractive sphere and rod gels. Attractive rods are able to form stable gels at much lower volume fractions compared to spheres as expected in the case of anisotropic particles. Moreover, linear viscoelasticity and the mechanism of yielding is significantly different in the two cases. Attractive sphere gels exhibit brittle nature and strengthen as pre-shear rate is reduced, whereas rod gels exhibit the reverse. To elucidate the structural contribution to this rheological response, we performed rheo-confocal studies on an attractive rod gel suspension and revealed that rod gels retain their isotropic nature under different pre-shear conditions. While the main contribution towards different viscoelastic response arises from changes in structural heterogeneity, it is also dependent on the type of pre-shear protocol.
	
	\section{\label{sec:level2}Materials and methods}
	\subsection{\label{sec:level2.1}Sample Preparation}
	
\begin{figure}[hbt!]
		\centering
		\includegraphics[width=1\linewidth]{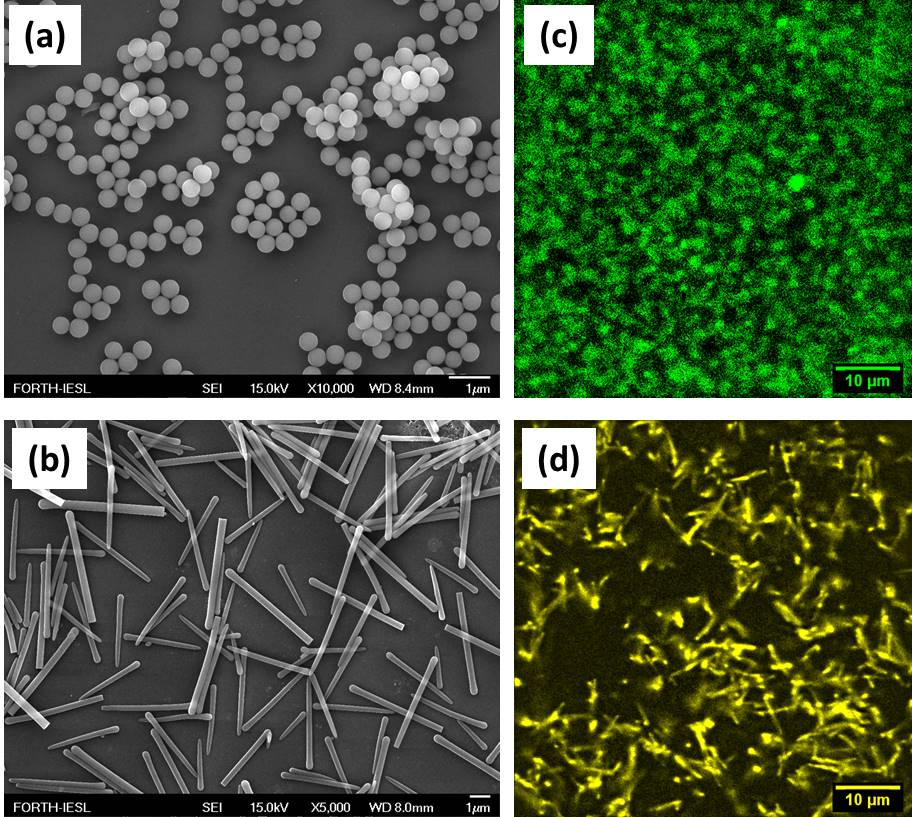}
		\caption{SEM images of silica spheres (a) and silica rods (b) used in the study and their corresponding 2D confocal micrographs (c \& d).}
		\label{fig:fig1}
	\end{figure}

	Monodisperse uncoated silica spheres were obtained from a commercial supplier (AngstromSphere) and silica rods were synthesized using well-established wet synthesis process.\cite{kuijk2011synthesis} Details regarding rod synthesis and particle density ($\rho$) measurements are provided in ESI. The density of spheres and rods were found to be almost the same, but not identical, i.e. $\rho_{sphere}$ = 1.95 and $\rho_{rod}$ = 1.98 g/cm$^3$ respectively. Particle dimensions were obtained from SEM images (Fig. \ref{fig:fig1}). Silica spheres have an average diameter, \textit{D} = 450 $\pm$ 0.03 nm and silica rods have an average length, \textit{L} = 4.8 $\pm$ 0.8 $\mu$m and \textit{L/D} = 10. Silica rods and spheres used for confocal microscopy were prepared by separating them from initial batch of particles and then fluorescently labeling them using Rhodamine B dye. 
	\par We prepared suspensions of silica spheres ($\phi$ = 0.20-0.40) and rods ($\phi$ = 0.02-0.10) by dispersing vacuum dried particles in 11 M CsCl aqueous solution first by stirring for 2 h using a magnetic pellet followed by sonication for 10 min. Particle volume fractions were chosen such that they form a gel in both cases and would have similar range of values for their elastic modulus. In the case of spheres the volume fractions were much above the Diffusion Limited Cluster Aggregation (DLCA) regime\cite{zaccarelli2007colloidal,mohraz2005orientation,guo2011gel} and above the regime where flow-induced instabilities such as vorticity aligned clustering (or \textit{log-rolling}) are observed.\cite{hoekstra2003flow,vermant2005flow,osuji2008highly,varga2019hydrodynamics} Similarly, in the case of rods the volume fractions chosen were $\phi\geq$ 0.04 where we do not observe any flow-induced instabilities or vorticity-aligned clusters as demonstrated in our previous study.\cite{das2020shear} 
	
	\subsection{\label{sec:level2.2}Interparticle attraction}
	The silica particles here are suspended in a 11 M CsCl aqueous solution. Due to the high salt concentration ($\kappa^{-1}<$ 1 nm), van der Waals attraction forces dominate. In the case of rods, the particle volume fraction is in the semi-dilute regime\cite{doi1988theory} (D/L)$^2\leq\phi\leq$(D/L) which implies that the suspension is isotropic in nature. Furthermore, in the case of charged rigid rods, their ability to form an ordered phase is counteracted by their tendency to aggregate perpendicular to each other. Stroobants et al.\cite{stroobants1986effect,drwenski2016phase,buining1994phase} showed that this effect is governed by a ``twist'' parameter, \textit{h = ($\kappa$D$_{eff}$)$^{-1}$}. This arises due to angular dependence of electrostatic interaction between charged rods and is more significant when the Debye length $\kappa^{-1}$ is of the order of the rod diameter. For spheres and crossed rods, the pair potential is governed by van der Waals forces at an interpaticle distance \textit{d $<<$ D} and is given by,\cite{israelachvili2015intermolecular}
	\begin{equation}
		U(d)=\frac{-A}{24}\frac{D_{sphere}}{d}
	\end{equation}
	\begin{equation}
		U(d)=\frac{-A}{12}\frac{D_{rod}}{d}
	\end{equation}
	The interparticle potential for attractive spheres and rods was found to be -21\textit{k$_B$T} and -45\textit{k$_B$T} respectively at an interparticle surface distance of 1 nm.  The range of attraction, $\delta$ was chosen where the interaction potential equals -1\textit{k$_B$T} and was found to be 0.045\textit{D$_{sphere}$} (26 nm) and 0.09\textit{D$_{rod}$} (49 nm) respectively. We assume that at shorter interparticle distances, the potential varies linearly with distance.
	\subsection{\label{sec:level2.3}Rheo-confocal measurements}
	All the rheological measurements were carried out using a stress controlled rheometer (MCR WESP 302, Anton Paar). We used home-made rough cone-plate geometry (made from polyether ether ketone to eliminate wall-slip) of diameter 40 mm, cone angle 2.5$^{\circ}$ and truncation 120 $\mu$m. Dodecane (Alfa Aeser, 99+ \%, with viscosity of 1.36 mPa s) was used as a solvent trap by pouring it around the sample after loading in the rheometer geometry. Samples were rejuvenated at high shear rate ($\dot\gamma$ = 1000 s$^{-1}$) before commencing any rheological measurement to erase sample loading and shear history. Rejuvenation was followed first by a dynamic time sweep (DTS) at $\gamma$ = 0.1 \%, $\omega$ = 1 rad s$^{-1}$ for 2000 seconds, then by dynamic frequency sweep (DFS) at $\gamma$ = 0.1 \%, $\omega$ = 0.1 - 100 rad s$^{-1}$ and finally by dynamic strain/amplitude sweep (DSS) at $\gamma$ = 0.1-1000 \%, $\omega$ = 1, 5 and 10 rad s$^{-1}$. It should be noted that DSS measurements at different $\omega$ were performed, after completing rejuvenation, DTS and DFS protocols. Flow curves for the suspensions were obtained by performing a high to low rate steady shear sweep and making sure to obtain steady state stress ($\sigma$) values for the shear rates applied. The steady state values of stress were determined by performing step rate measurements.
	\par Flow cessation experiments were carried out to determine time evolution of viscoelasticity (aging) after the gel stops flowing. This was repeated for different pre-shear rates under steady and oscillatory shear flow. Here we first performed high shear rejuvenation ($\dot\gamma$ = 1000 s$^{-1}$) for 300 seconds, stopping for 3 seconds ($\dot\gamma$ = 0) and then performing either steady shear flow at constant shear rates ($<$ 1000 s$^{-1}$) or large amplitude oscillatory shear (LAOS) measurements ($\gamma \ge$ 10 \%, $\omega$ = 10 rad s$^{-1}$) for 600 seconds which forms the ``pre-shear'' protocol. This was followed by DTS, DFS and DSS ($\omega$ = 1 rad s$^{-1}$) measurements. As mentioned earlier, due to the interparticle attraction of van der Waals type, shear rates were normalized by a dimensionless Mason number, Mn = \textit{F$_{Visc.}$/F$_{Attr.}$ = 6$\pi \eta_{0} a^{2}\delta\dot{\gamma}$/U} where F$_{Visc.}$ is the viscous drag force on particle of radius \textit{a} and \textit{F$_{Attr.}$ = U/$\delta$} is the interparticle attraction force. At Mn $>$ 1, interparticle attraction force can be overcome by viscous drag force leading to particle dispersion. Shear rates in steady ($\dot\gamma$) and oscillatory ($\gamma_0 \omega$) pre-shear measurements were chosen such that they have the same dimensionless shear, denoted by Mn for steady shear and Mn$_\omega$ for oscillatory shear.
	\par A confocal microscope (VT-Eye, Visitech International) was attached to the rheometer using an in-house setup to capture the gel microstructure during shear flow. For this we replaced the rough plastic bottom plate with a steel plate with a glass coverslip (thickness = 170 $\mu$m) glued on top. The coverslip surface was roughened by depositing a uniform thin layer of crushed glass powder with average particle size of 50 - 80 $\mu$m. We used Nikon oil immersed 100 $\times$ objective (NA = 1.45, working distance = 130 $\mu$m) mounted on a piezoelectric stage for fast z-scans. Above the objective, a transparent window large enough to capture confocal micrographs of size 50$\times$50 $\mu$m$^2$ was created by removing the glass powder from the top of the coverslip. Rheological results from regular geometry were compared with rheo-confocal setup to ensure reproducibility. 
	\section{\label{level3}Results}
	\subsection{\label{level3.1}Steady shear response and yield stress}
\begin{figure}[hbt!]
		\centering
		\includegraphics[width=1\linewidth]{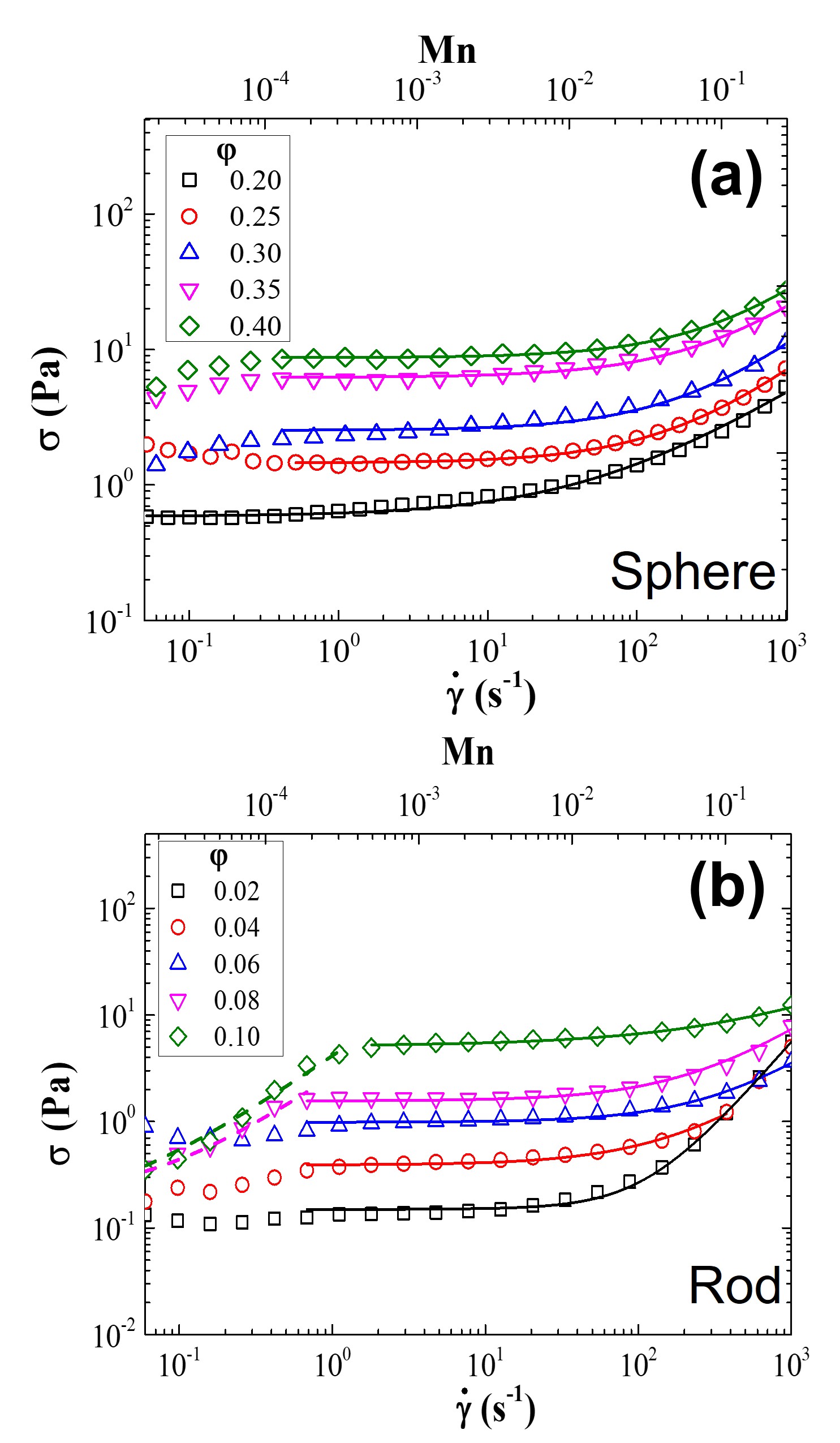}
		\caption{Flow curves of gels of attractive spheres (a) and rods (b). Solid lines are Herschel-Bulkley fits. Stronger gels exhibit wall slip at the lowest shear rates depicted here by a drop in steady-state stress values. Dotted line are Bingham model fits to determine slip stress ($\sigma_s$).}
		\label{fig:fig2}
	\end{figure}

	We measured the flow curve of the colloidal gels after rejuvenation by performing a reverse constant shear rate sweep and measuring the stress response. We observe that both attractive sphere and rod gels exhibit shear thinning behaviour and at high shear rates the stress increases linearly with shear rate indicating a viscous flow behaviour (Fig. \ref{fig:fig2}). We fit the data in both cases using Herschel-Bulkley model (HB), \textit{$\sigma$ = $\sigma_y$ + k$\dot{\gamma}^n$}. The value of the exponent \textit{n} in the HB fits for flow curve of gel of attractive rods is less than 1 indicating shear thinning at the highest shear rates. Shear thinning behaviour of both type of gels is in agreement with the fact that even at the highest shear rate applied in the rheometer (\textit{Mn} = 0.3), it is impossible to break all the aggregates into individual particles. At lower shear rates, we observe a drop in shear stress values in both gels, which is indicative of wall-slip. However, this drop is stronger in gels of attractive rods and occurs at higher shear rates (or Mn) compared to gels of attractive spheres. Such drop leads to a second stress plateau where a \textit{slip stress} can be determined. We should also note that for rod gels ($\phi \leq$ 0.04) we observe a mild shear thickening at high rates as the stress is increasing stronger than linearly with shear rate.
	\par In general, colloidal gels exhibit wall-slip at low shear rates and this problem can be countered by using a rough geometry.\cite{ballesta2013slip} However, the degree of roughness is also important in order to suppress wall-slip in colloidal gels. As the shear rate is reduced, Mn decreases which means the viscous drag forces become weaker than interparticle bonds. This leads to an increased particle clustering and heterogeneity in the gel microstructure\cite{koumakis2015tuning} and therefore leads to less number of contact points between the particle network and the wall. In addition, it has been demonstrated through experiments and simulations that under no-slip condition, low shear rates (or \textit{Mn}$<<$1) promote cluster densification to allow for easier flow.\cite{torres1991floc,potanin1993computer,mohraz2005orientation,koumakis2011two,laurati2011nonlinear} While designing rough surfaces, it has to be kept in mind that the dimensions of roughness should scale approximately with the size of largest particle clusters present. Since it is almost impossible to determine the precise size of a cluster when using non-transparent geometries, a safe method would be to introduce roughess of the order of 50-100 particle length. A similar value for cone truncation needs to be considered to avoid confinement effects.\cite{ramaswamy2017confinement}
	
	\par Understanding wall-slip phenomena when comparing gels of attractive spheres and rods is more complicated as particle cluster size, structural heterogeneity, cluster densification and the magnitude of roughness with respect to the cluster size all play a role in promoting wall-slip in attractive particle suspensions. Under isotropic conditions; the random close packing limit of rods of length \textit{L} and diameter \textit{D} is given by, $\phi_{rcp}$(\textit{L/D}) $\approx$ 5.4 ($\pm$0.2).\cite{philipse1996random} For the rods used here with \textit{L/D} = 10,  $\phi_{rcp}$ = 0.54 which is lower compared to that of spheres $\approx$ 0.64. Moreover, rods can form larger and less dense clusters compared to spheres due to their higher excluded volume.\cite{mohraz2004effect} In low volume fraction isotropic gels of attractive rods, the number of contact points between rods and the wall will be lower compared to a high volume fraction gel of attractive spheres. Hence, it can be expected that gels of attractive rods would exhibit more prominent wall-slip compared to that of attractive spheres. Moreover, if there is rod alignment close to the wall resulting from bond breaking between rod pairs at \textit{Mn} $>$ 1, the mechanism of wall-slip may be different than above.
	
	\par Yield stress ($\sigma_y$) for both systems were determined from Herschel Bulkley fit to the flow curve data. The fits were performed for shear rates above which the no-slip criteria was satisfied. Values of \textit{slip stress} ($\sigma_s$) in the low shear rate regime (indicated by dotted curve in Fig. 2 (b)) of high volume fraction gels of attractive rods were obtained using a Bingham model fit given by $\sigma = \sigma_s + \eta\dot{\gamma}$,\cite{mewis2012colloidal}
	
\subsection{\label{level3.2}Aging and linear viscoelasticity}
Here we follow the linear rheological response of the attractive silica sphere and rod suspensions after subjecting them to high shear rejuvenation. The rejuvenation step was followed by a DTS measurement where both attractive sphere and rod gels exhibit time evolution of their viscoelasticity (aging). The increase in the gel elasticity can be attributed to repercolation of the particle network after breakage during rejuvenation. However, we do not observe a distinct transition in time from a shear melted liquid to yield stress solid after rejuvenation for either gel as had been observed at lower volume fraction ($\phi \le$ 0.008) rod gels in our pervious study.\cite{das2020shear} Rheo-confocal results from the same study revealed that at higher volume fraction or $\phi \ge$ 0.04 ($\phi/\phi^*$ = 5), although there were no obvious changes in rod gel microstructure, a time-dependent increase in gel elasticity from a weaker to a stronger solid was observed. We claimed that this is due to the dense nature of the gel network and that aging in gels formed by screening Coulomb interactions is driven by stiffening of solid-solid particle contact points known as contact-controlled aging.\cite{bonacci2020contact} A similar effect is observed with our results here as well, where the gel structure repercolates quickly after rejuvenation ($<$ 6 seconds) and thereafter strengthens over time. 
\par It is important to note that the highest shear rate applied for rejuvenation (1000 s$^{-1}$) corresponds to Mn = 0.3 and hence we do not expect to fully break interparticle bonds. The repercolation after rejuvenation would therefore be due to reformation of inter-cluster bonding of smaller clusters formed as a result of rejuvenation.
\begin{figure}[hbt!]
	\centering
	\includegraphics[width=1\linewidth]{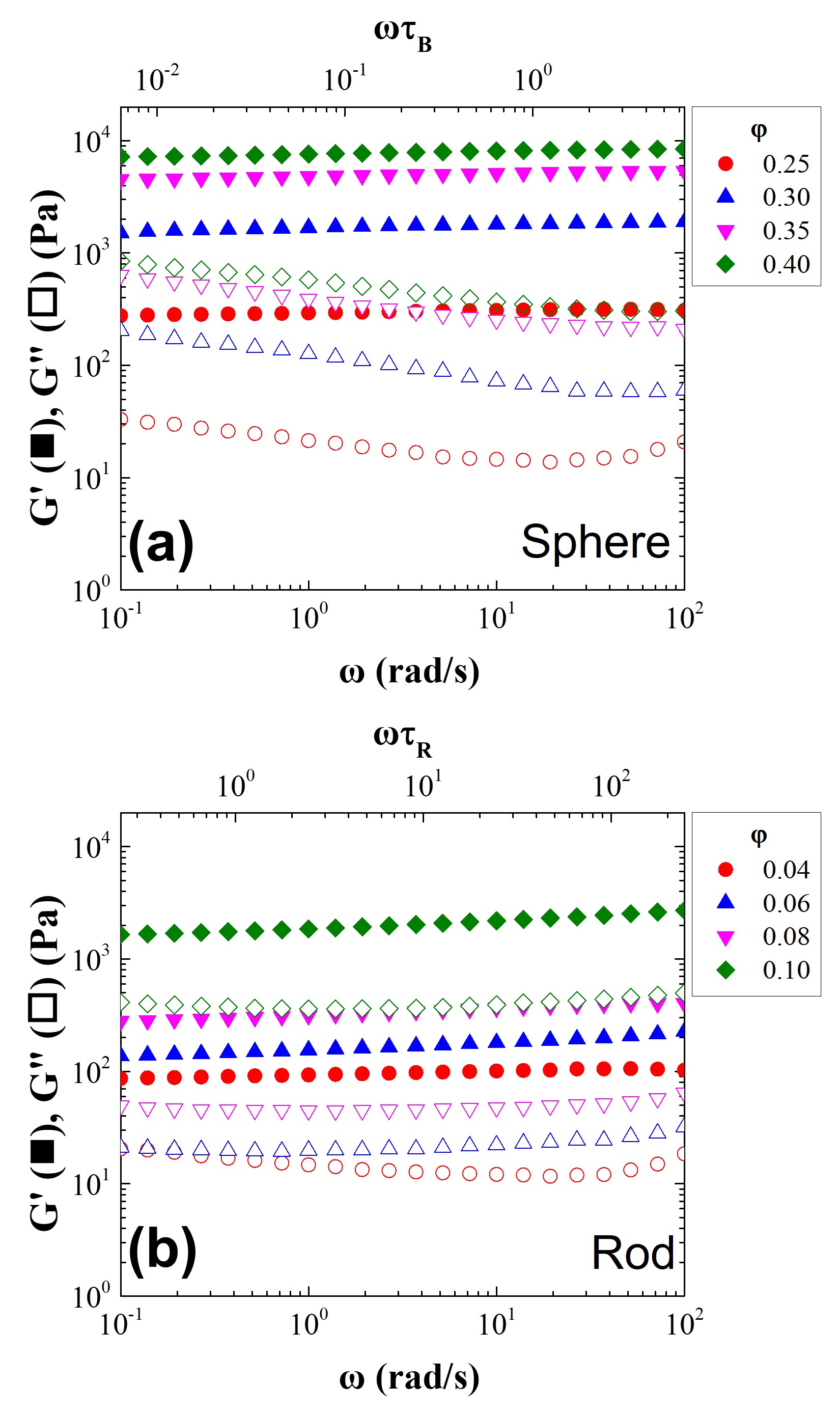}
	\caption{DFS measurements after t$_w$ = 2000 s of attractive sphere (a) and rod (b) gels comparing structural relaxation at different time scales measured at $\gamma$ = 0.1 \%. Here $\tau_B$ = R$^2$/D$_t$ for spheres and $\tau_r$ = 1/6D$_r$ for rods.}
	\label{fig:fig3}
\end{figure}
\par DFS measurements performed after DTS show that the suspensions exhibit solid-like behaviour with a non-terminal flow response detected at low frequencies. We normalized the angular frequency here by the Brownian time, $\tau_B$ (= \textit{R$^2$/D$_t$}) for spheres and rotational relaxation time, $\tau_r$ (= \textit{1/6D$_r$}) for rods. In both cases, G' being almost frequency independent indicates the formation of a strong gel where particle level and cluster level diffusion is completely arrested.

\subsection{\label{level3.3}Nonlinear dynamic response}
\begin{figure}[hbt!]
	\centering
	\includegraphics[width=1\linewidth]{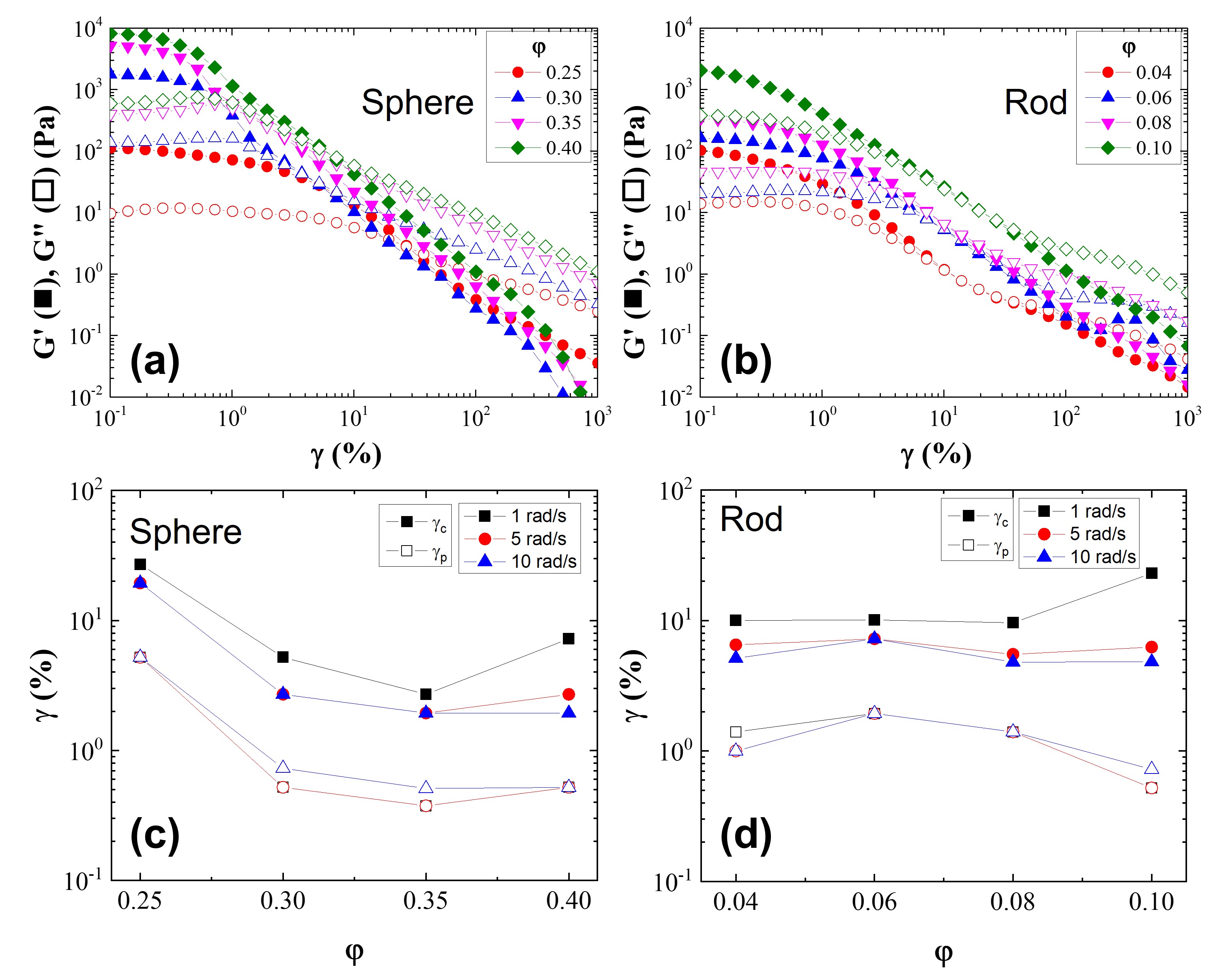}
	\caption{DSS measurement performed at $\omega$ = 1 rad/s for gels of attractive spheres (a) and rods (b) at different particle volume fractions ($\phi$) and their corresponding $\gamma_{c}$ and $\gamma_{p}$ (c) \& (d). Lines are drawn in (c) \& (d) to guide the eye.}
	\label{fig:fig4}
\end{figure}

\begin{figure}[hbt!]
	\centering
	\includegraphics[width=1\linewidth]{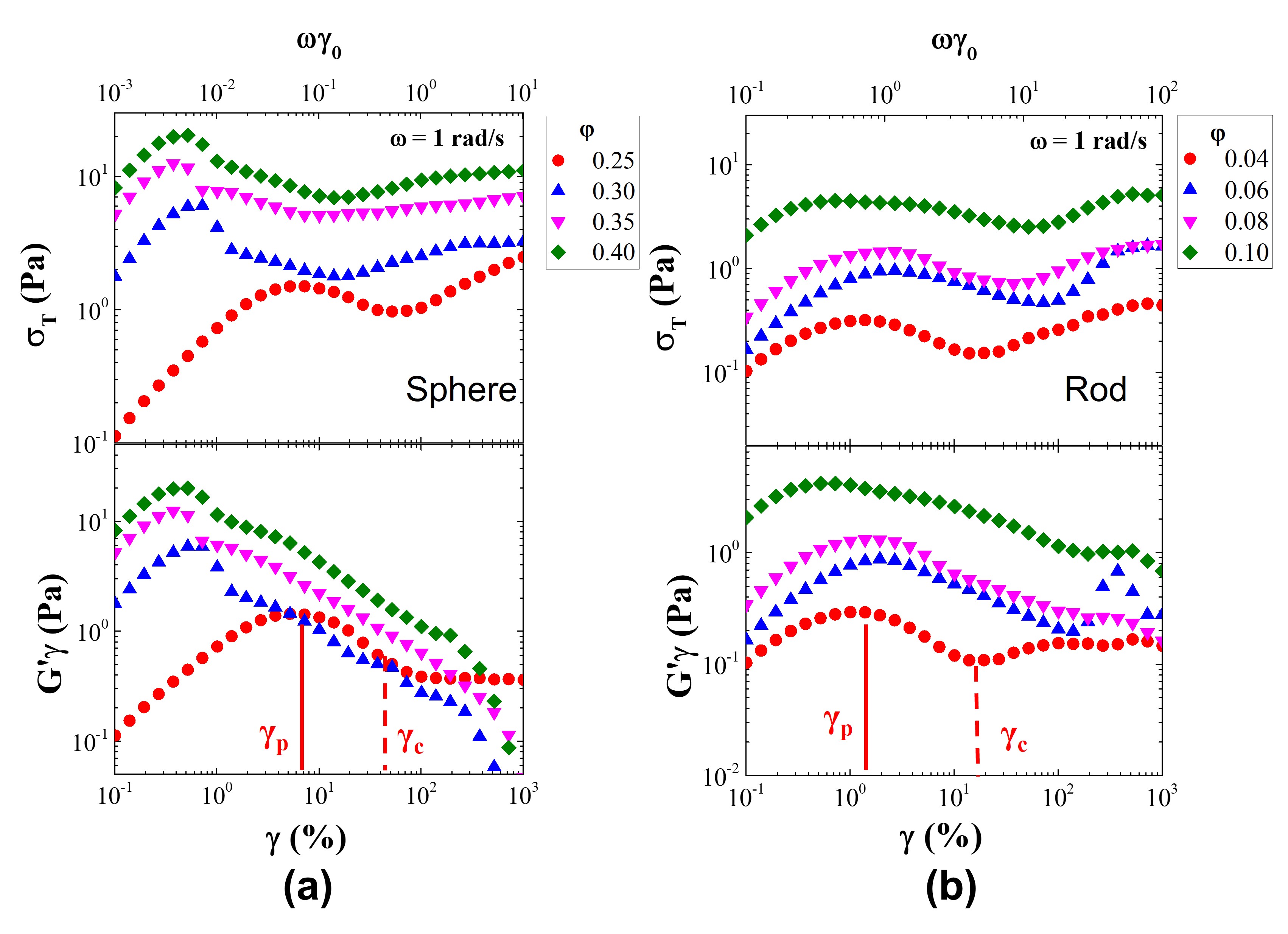}
	\caption{Total stress ($\sigma_T$) and elastic stress (G'$\gamma$) obtained from dynamic strain sweep measurement at $\omega$ = 1 rad s$^{-1}$ plotted as a function of strain for gels of attractive spheres (a) and rods (b). The maximum in elastic stress curve is defined as the yield stress of the LAOS measurement. Solid line - $\gamma_{p}$, dotted line - $\gamma_{c}$.}
	\label{fig:fig5}
\end{figure}

We probe the dynamic yielding behaviour of the gels by performing DSS measurements at $\omega$ = 1, 5 and 10 rad s$^{-1}$ from $\gamma$ = 0.1 to 1000 \%, after DTS (t$_w$ = 2000 seconds) and DFS. Yielding of both types of gel seems almost independent of $\omega$ except at very high $\gamma$ ($>$ 100\%). Moreover, a sharp drop in G' is observed at very large $\gamma$ which could be due to some wall-slip which increases with $\omega$. We observed that this behaviour is stronger in gels of attractive spheres than rods, especially at $\omega$ = 1 rad s$^{-1}$. The second observation is regarding yielding response. The yield point can be identified either by the crossover strain, where G'= G'', denoted here as $\gamma_{c}$ or the peak of the G'' that signifies the maximum energy dissipation, denoted here as $\gamma_{p}$ during yielding of gels at different particle volume fractions as shown in Fig. \ref{fig:fig4}. Here we see that $\gamma_{c}$ is reduced as $\omega$ is increased (Fig.\ref{fig:fig4} (c) and (d) whereas this effect is less prominent for $\gamma_{p}$. A third finding is that attractive sphere gels become more brittle with increasing volume fraction compared to attractive rod gels since both $\gamma_{c}$ and $\gamma_{p}$ for spherical particle gels decrease with increasing volume fraction, whereas they are almost $\phi$-independent for rod gels. Note that $\gamma_{c}$ and $\gamma_{p}$ for gels of spheres here are smaller compared to that of hard-sphere depletion gels formed by colloid-polymer mixtures.\cite{mewis2012colloidal,laurati2011nonlinear} In addition, we do not observe any clear sign of two-step yielding as has been demonstrated in some previous studies.\cite{laurati2011nonlinear,koumakis2011two,chan2012two} These observations point towards important differences in the mechanism of yielding in colloidal gels formed by particles with strong van der Waals attraction forces with short range. Moreover, attractive sphere gels studied here are denser compared to rod gels and would therefore form stronger particle networks. We finally note that the linear regime for spherical particle gels extends up to about $\gamma \approx$ 5\% whereas rod gels indicate a less clear G' plateau at low strains, and therefore a narrower linear regime, indicative probably of a higher structural network flexibility.
Further, we also determined the yield stress of the material from the dynamic strain sweep as the peak (maximum) of the elastic stress, G'$\gamma$, as has been used previously in different systems.\cite{walls2003yield,yang1986some,pai2002gelation,koumakis2011two,laurati2011nonlinear} As shown in Fig.\ref{fig:fig5}, the total stress ($\sigma_T$) and the elastic stress (G'$\gamma$) measured during dynamic strain sweep measurements at $\omega$ = 1 rad s$^{-1}$ maybe used to identify a dynamic yield stress. Here a linear increase in stress with strain indicates the elastic response of the gel network. At higher strain amplitudes the gel yields, the stress drops and the system exhibits a viscous flow with $\sigma_{T}$ (or G'$\gamma) \propto$ $\dot{\gamma}$ $\sim$ $\omega\gamma_0$. Difference in the broadness of stress peaks in both type of gels further reveals a more sharp yielding of spherical gels in comparison with a rather more gradual  yielding of rod gels. 

\begin{figure}[hbt!]
	\centering
	\includegraphics[width=0.7\linewidth]{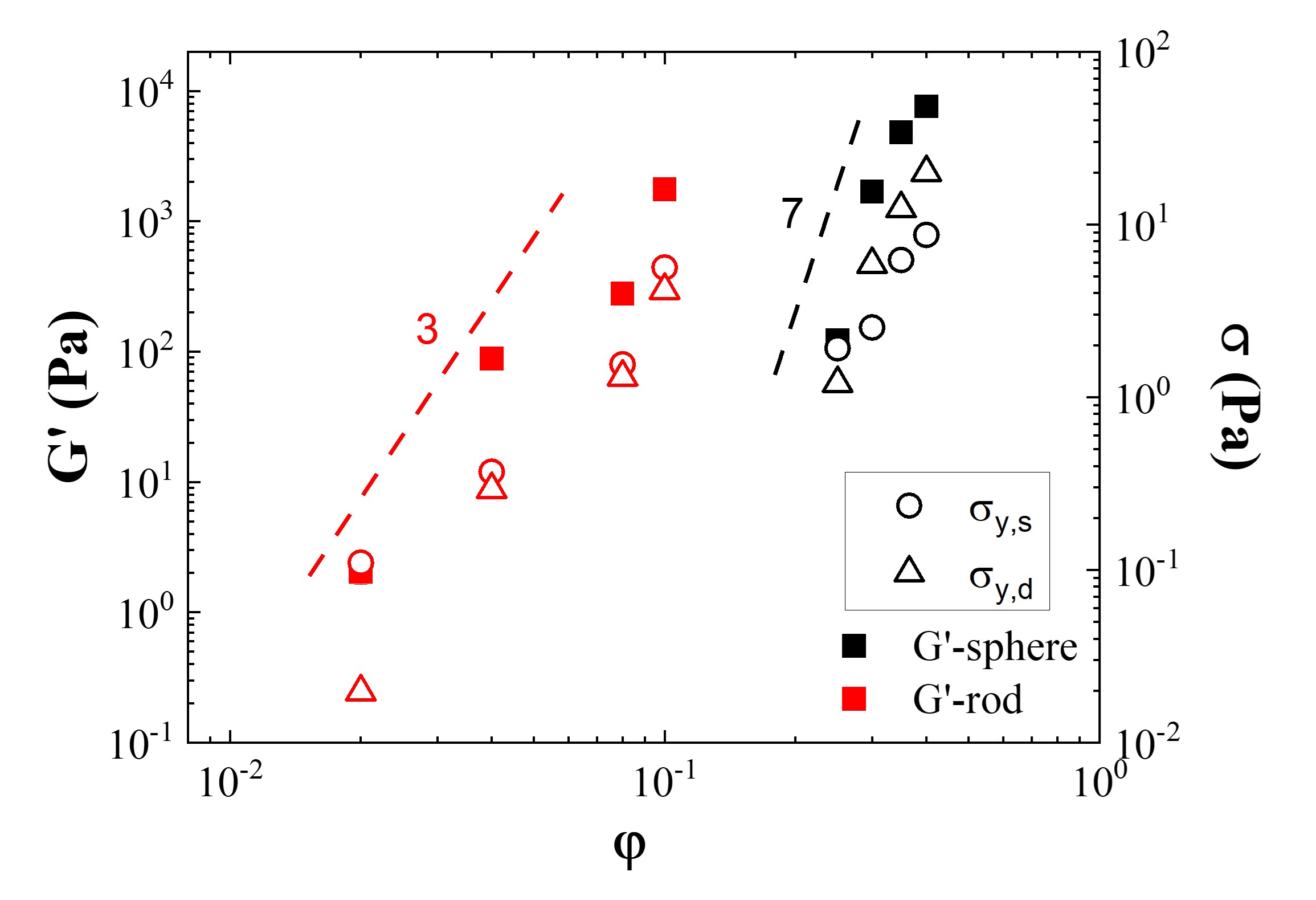}
	\caption{Storage modulus (G' at $\gamma$ = 0.1\% and $\omega$ = 1 rad/s) after t$_w$ = 2000 seconds from rejuvenation and yield stress obtained using flow curve ($\sigma_{y,s}$) and dynamic strain sweep ($\sigma_{y,d}$) plotted as a function of particle volume fraction $\phi$. Dotted line are approximate power-law fits for G' vs. $\phi$.}
	\label{fig:fig6}
\end{figure}
\par Values of yield stress obtained using both methods (dynamic strain sweeps and flow curves) are plotted in Fig. \ref{fig:fig6}. Here we observe that for gels of attractive spheres, the values of $\sigma_y$ obtained using Herschel Bulkley fit (denoted $\sigma_{y,s}$) are lower compared to that from DSS measurements (denoted $\sigma_{y,d}$). However, for gels of attractive rods the values were comparable. It is important to understand that the type of flow field employed in both type of measurements is different and colloidal gel structure is sensitive to such changes.\cite{koumakis2015tuning,moghimi2017colloidal} Hence, some level of discrepancy may be expected.
Comparing the volume fraction dependence of G' and $\sigma_{y}$ measured from steady and oscillatory shear measurements show that attractive sphere gels exhibit a stronger volume fraction dependence of G' ($\propto \phi^7$, approx.) compared to attractive rod gels ($\propto \phi^3$, approx.). Note, however, in this limited $\phi$ range, the approximate value of power law exponent for gels of attractive spheres is higher than the values reported for depletion gels.\cite{krall1998internal,romer2014rheology,kao2022microstructure} It needs to be verified if this is a result of contact driven aging observed in colloidal gels formed by interparticle van der Waals attraction.\cite{pantina2005elasticity,pantina2006colloidal,bonacci2020contact}
\subsection{\label{level3.4}Response to pre-shear}
We now discuss the role of shear history (steady vs. oscillatory) on the linear viscoelastic response and yielding of gels of attractive spheres and rods. We subjected the colloidal gels to different pre-shear protocols (steady and oscillatory) after rejuvenation (1000 s$^{-1}$) until steady state was reached and then measured the linear (DTS and DFS) and nonlinear (DSS) viscoelastic response. While applying pre-shear, we ensured that the range of shear rates in both type of flows ($\dot{\gamma}$ and $\gamma_0 \omega$) were the same. From here on, we denote Mason numbers related to steady shear as Mn and oscillatory shear as Mn$_\omega$. Shear rates at which wall-slip was encountered during flow curve measurements were not considered. 

\par
\begin{figure}[hbt!]
	\centering
	\includegraphics[width=1\linewidth]{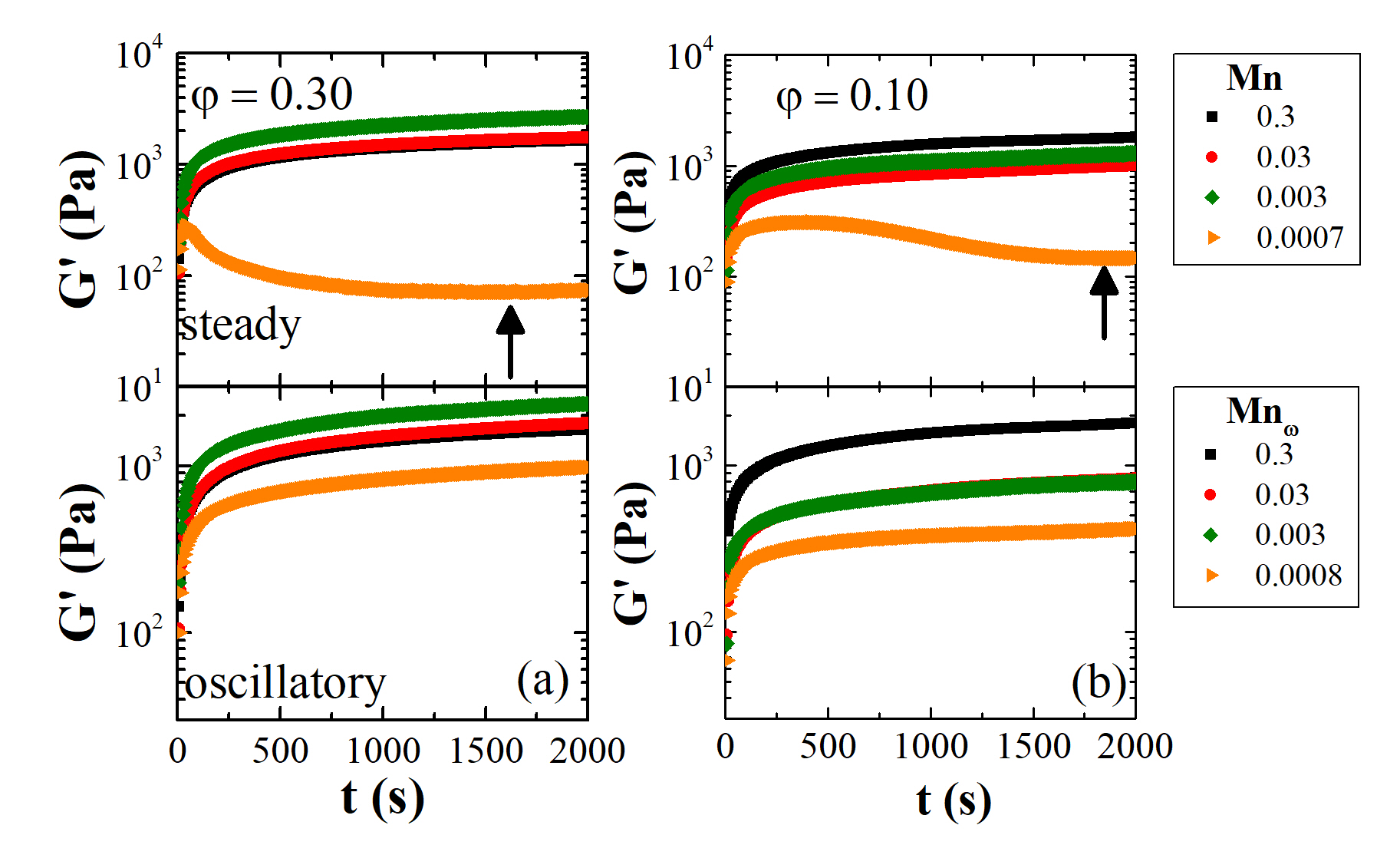}
	\caption{Evolution of G' after pre-shear at different \textit{Mn} under steady (top) and oscillatory (bottom) shear flow for gels of  (a) attractive silica spheres ($\phi$ = 0.30) and (b) attractive silica rods ($\phi$ = 0.10). Arrow indicates probable sedimentation.}
	\label{fig:fig7}
\end{figure}

 We first looked at the aging of gels of both attractive spheres and rods after subjecting them to different types of pre-shear (Fig. \ref{fig:fig7}). In the case of spheres, we observe that all the samples exhibit solid-like behaviour which strengthens with time after shear cessation. However, after very low steady pre-shear; we observe a drop in G' with time (shown by arrow in Fig. \ref{fig:fig7} (a) - Top). In contrast, this effect was not so prominent after oscillatory pre-shear and was observed only in one of the gel samples. Such a strong drop in elasticity of the gel represents a break up of the gel network and possible sedimentation. Note that due to poor density matching between the particles and solvent, at very long waiting times (beyond experimental time) the gels would eventually sediment. For gels of  attractive rods (Fig. \ref{fig:fig7} (b)), we observed a drop in G' during aging only for the highest volume fraction gel ($\phi$ = 0.10), again after low steady pre-shear, whereas in all other cases the gel was stable during the aging measurement. 

\par
\begin{figure}[hbt!]
	\centering
	\includegraphics[width=1\linewidth]{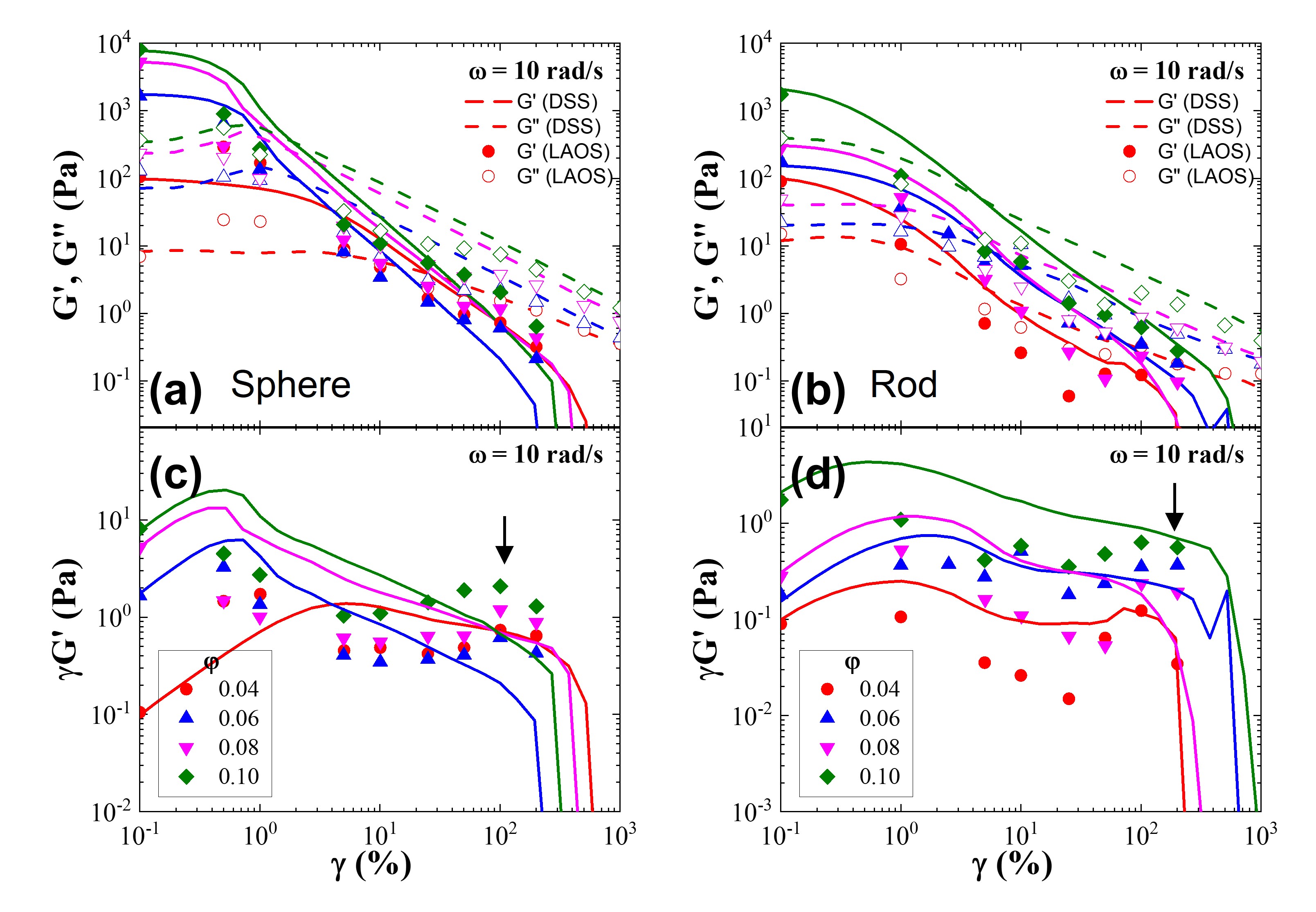}
	\caption{Comparison of G' and G'' and elastic stress (G'$\gamma$) values obtained from DSS measurements (lines) and LAOS measurements (symbols) at $\omega$ = 10 rad s$^{-1}$ and taken after t$_w$ = 600 seconds for gels of attractive spheres and rods. Arrows indicate $\gamma$ where a second yielding process is expected.}
	\label{fig:fig8}
\end{figure}

 Next we compare the steady state values of G' and G'' obtained after t$_w$ = 600 s during large amplitude oscillatory pre-shear (LAOS) of the gels with that obtained during DSS measurements (Fig. \ref{fig:fig8} (a) \& (b)). Steady state values of G' and G'' obtained during oscillatory pre-shear (in DTS) are lower than the values obtained during DSS measurement. Significant difference in results between the two measurements arises in the nonlinear regime ($\gamma > \gamma_{y}$). In the intermediate strain amplitude regime (10\% $< \gamma <$ 300\%) the values of both G' and G'' show a mild increase before dropping with increasing strain amplitude indicating a ``second'' yielding process. In order to better understand this process, we looked at changes in the elastic stress (G'$\gamma$) with strain amplitude (Fig. \ref{fig:fig8} (c) \& (d)) and observed a clear second peak in stress at intermediate strain amplitudes which indicates that there is indeed a two-step yielding process in both type of gels when subjected to oscillatory pre-shear. Hence, we note here that the values of G' and G'' obtained from DSS measurement are different from that of DTS measurement in the LAOS regime for gels. This is because during DSS measurement, the values of G' and G'' obtained are not necessarily steady state as that might be reached at much longer waiting times. However, in the case of colloidal gels that undergo structural evolution under LAOS ($\gamma \ge \gamma_{y}$), one needs to wait until steady state values of G' and G'' is reached. Hence, G' and G'' values obtained from DTS measurements in LAOS regime are the true response of the gel.

\par
\begin{figure}[hbt!]
	\centering
	\includegraphics[width=1\linewidth]{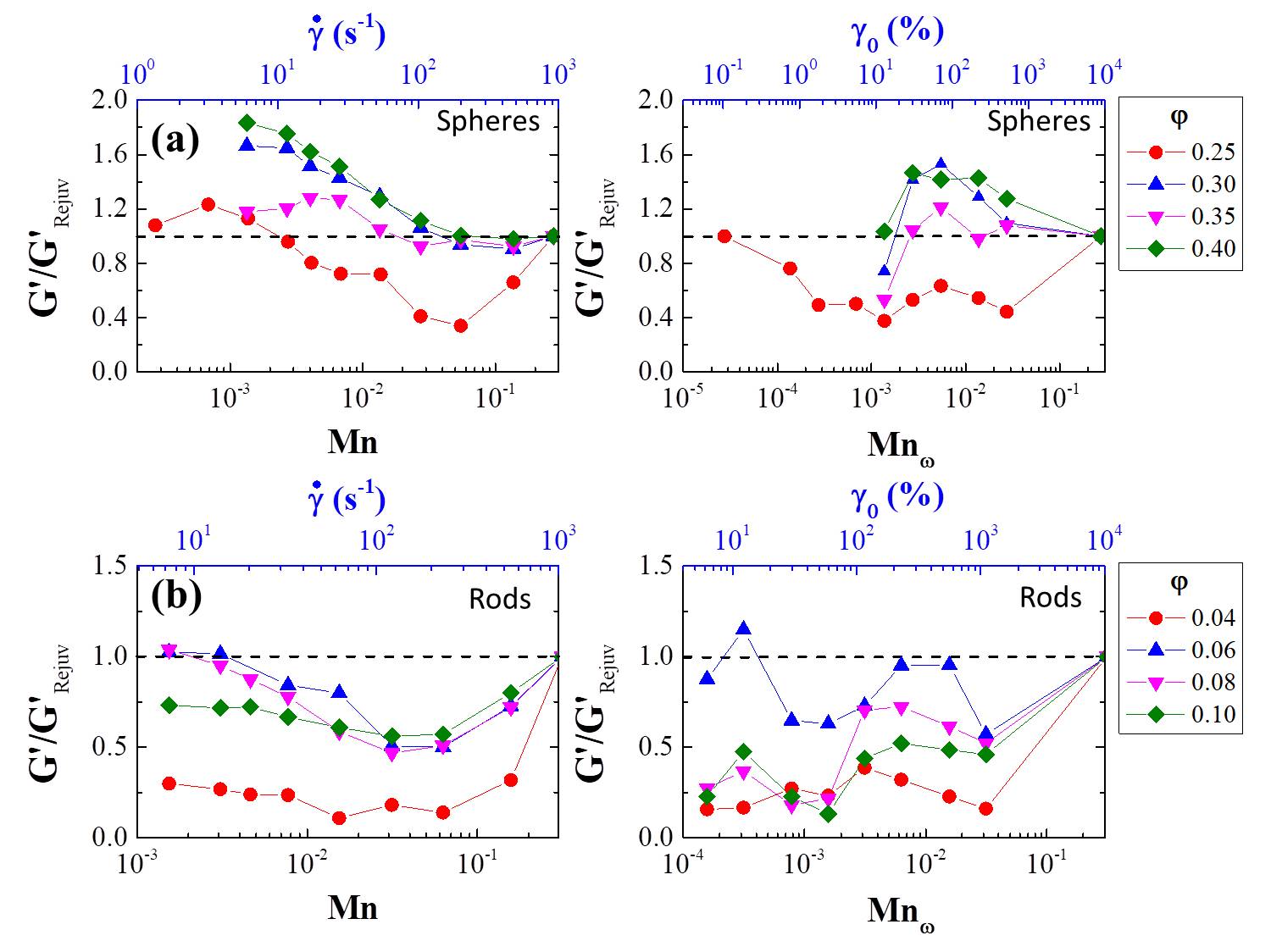}
	\caption{G'(at $\gamma$ = 0.1 \%, $\omega$ = 1 rad s$^{-1}$) obtained after aging (t$_w$ = 2000 s) for gels of attractive spheres (a) and rods (b) that underwent steady (left) and oscillatory (right) pre-shear. G' here is normalized by G' of gel after rejuvenation (G'$_{Rejuv}$) and represents its tunability.}
	\label{fig:fig9}
\end{figure}

\begin{figure}[hbt!]
	\centering
	\includegraphics[width=1\linewidth]{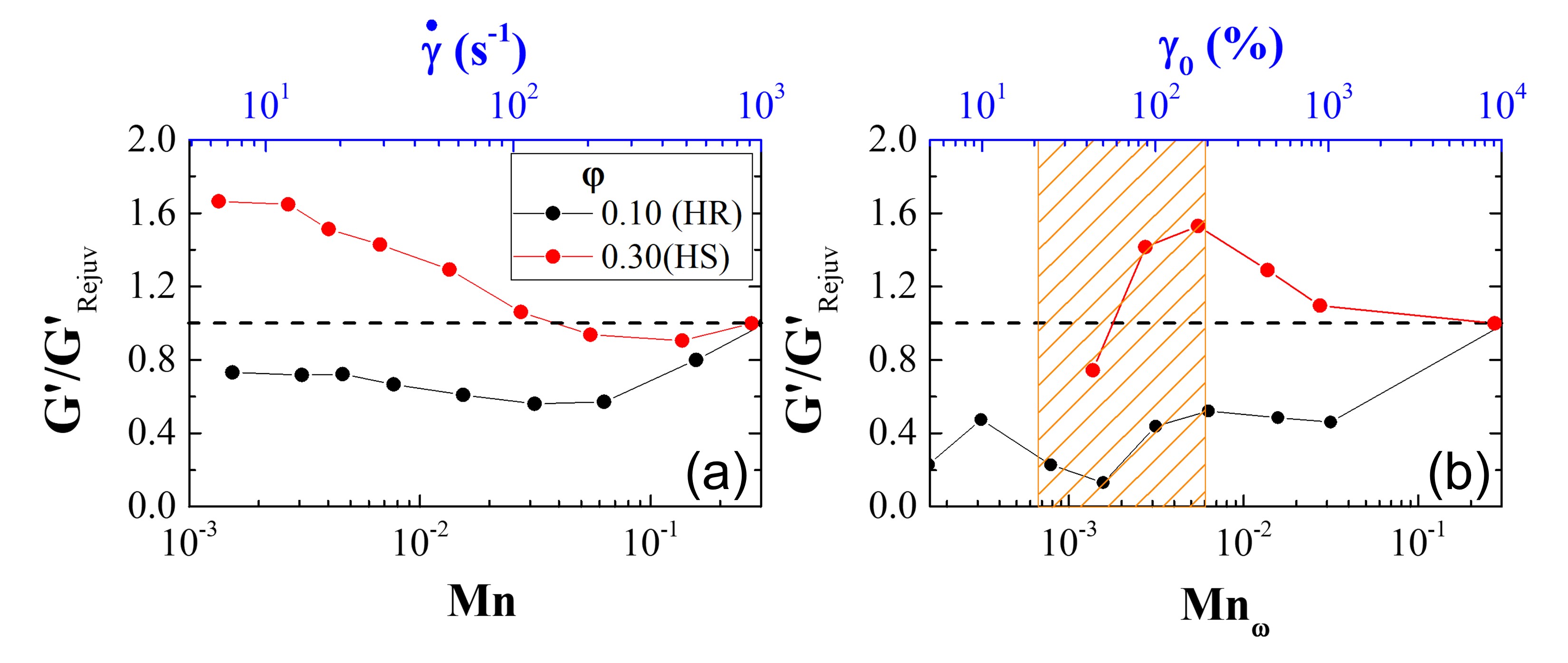}
	\caption{Tunability compared for gels of attractive spheres and rods of similar storage modulus after steady (a) and oscillatory (b) pre-shear. Here G'(at $\gamma$ = 0.1 \%, $\omega$ = 1 rad s$^{-1}$) obtained after aging (t$_w$ = 2000 s) following different pre-shear is normalized by G' of gel after rejuvenation (G'$_{Rejuv}$). Shaded region in (b) denotes strain amplitudes at which we expect sedimentation.}
	\label{fig:fig10}
\end{figure}

 We now examine the influence of pre-shear on the elastic modulus G' (or solid-like response) of gels of attractive spheres and rods. For this we normalize the values of G' after different pre-shear protocols, by value of G' obtained after rejuvenation (denoted here as G'$_{rejuv}$). G' values (at $\gamma$ = 0.1 \%, $\omega$ = 1 rad s$^{-1}$) were chosen from DFS measurements performed after t$_w$ = 2000 seconds after pre-shear. Pre-shear rates at which we observed sedimentation of gel were not considered for this analysis. In the case of gels of attractive spheres, at $\phi >$ 0.25 (Fig. \ref{fig:fig9} (a)) we see that as the steady pre-shear rate (denoted Mn) is reduced, the resultant gel structure becomes stronger than at quiescent state (G' $>$ G'$_{rejuv}$) with the values close to two times higher for the lowest shear rates applied. We see a similar strengthening of the gel after oscillatory pre-shear in the intermediate strain regime (50\% $<\gamma_0 <$ 200\%). However, as the strain amplitude of oscillatory pre-shear is reduced further, G' drops sharply. Note here that for oscillatory pre-shear we did not plot G' for $\gamma_0 < \gamma_{cross}$ for $\phi >$ 0.25 due to sedimentation of the resultant gel structure. In contrast, for gels of attractive rods; a reduction in pre-shear rate leads to weakening of the gels after both steady as well as oscillatory pre-shear (Fig. \ref{fig:fig9} (b)). Moreover, the weakening effect is stronger after oscillatory pre-shear. Similar to gels of attractive spheres, gels of attractive rods show a mild strengthening at intermediate strain amplitudes (50\% $< \gamma <$ 200\%) before undergoing further weakening. For better comparison between both type of gels, we compare the data of gels of attractive spheres and rods of similar strength at quiescent state in Fig. \ref{fig:fig10}. Here the contrast between the effect of pre-shear on both type of gels can be observed clearly. Gels of attractive spheres strengthen whereas that of rods weaken with reduction in Mn. For both gels, oscillatory pre-shear at intermediate strain amplitudes leads to strengthening, the effect of which is higher in gels of attractive spheres. 
\par
\begin{figure}[hbt!]
	\centering
	\includegraphics[width=1\linewidth]{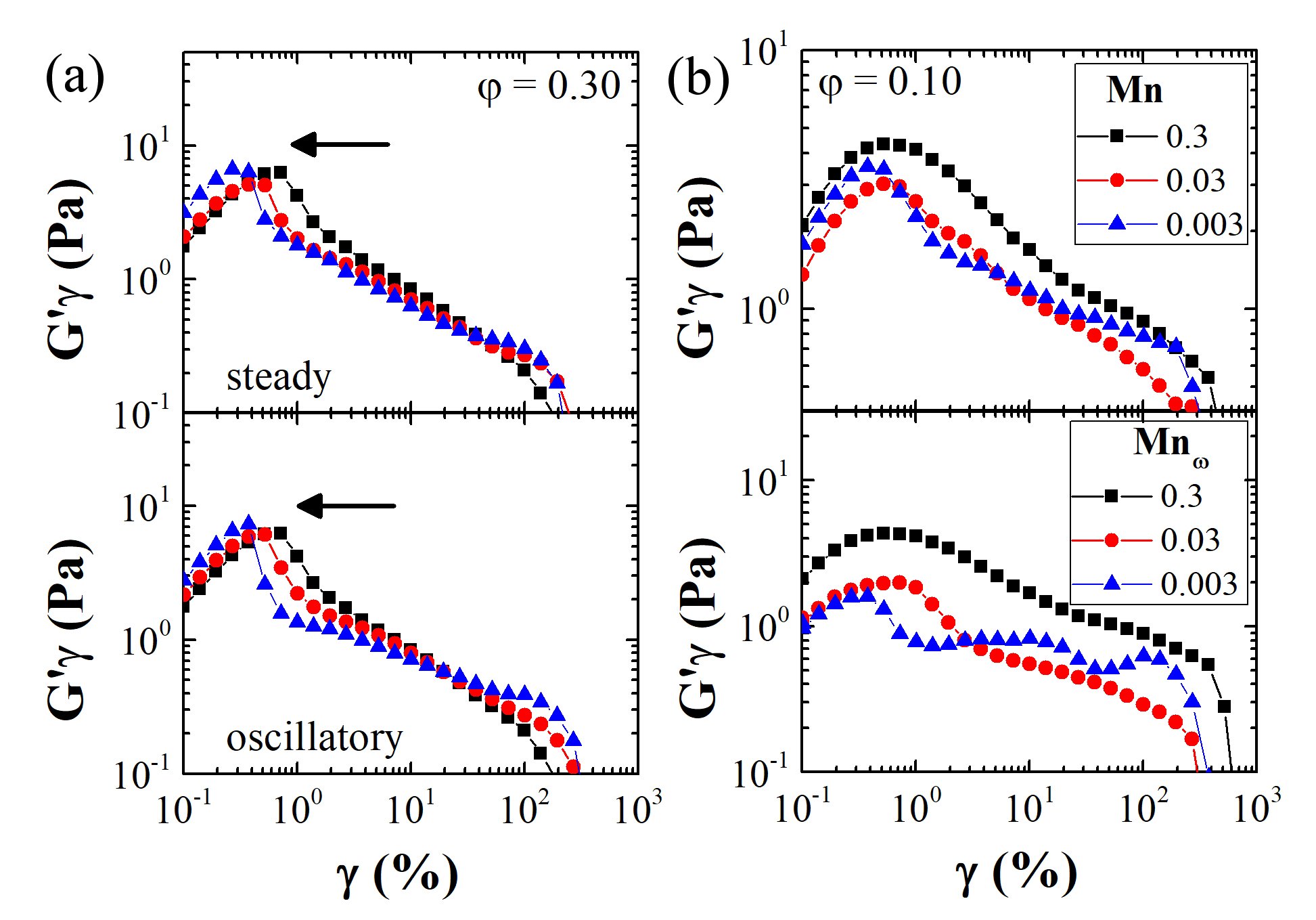}
	\caption{Elastic stress ($\gamma$G') plotted as a function of strain amplitude for gels of similar G' formed by attractive spheres (a) and attractive rods (b) pre-sheared under steady (top) and oscillatory (bottom) shear for $\phi_{sph}$ = 0.30 and $\phi_{rod}$ = 0.10. Arrows indicate the shift in position of elastic stress peak.}
	\label{fig:fig11}
\end{figure}

 We next examine the yielding behaviour of the new gel structure created as a result of different pre-shear protocols. For this we performed DSS measurements at $\omega$ = 10 rad s$^{-1}$. In the case of gels of attractive spheres we observe that with decreasing pre-shear rate (Mn and Mn$_\omega$) the elastic stress peak shifts to lower strain amplitudes (Fig. \ref{fig:fig11}, arrow indicates direction of peak shift). This behaviour was independent of the type of pre-shear and indicated the increasing brittleness of the gel with reducing pre-shear rates. However, gels of attractive rods did not exhibit such a clear trend. Moreover, we observe large fluctuations in the elastic stress at $\gamma \approx$ 100 \%. Additional information regarding structure-rheology relationship will be shown in the following section.
\par We can now compare our results with previous studies on tuning of hard-sphere colloid-polymer depletion gels using pre-shear.\cite{koumakis2015tuning,moghimi2017colloidal} Our results are in contrast to depletion gels where it was shown that a reduction in steady or oscillatory pre-shear rate leads to significant weakening of the gel. Moreover, it was reported by Moghimi \textit{et al.}\cite{moghimi2017colloidal} that oscillatory pre-shear at intermediate strain amplitudes leads to much weaker gels due to increased structural heterogeneity arising from densification of the particle clusters. In order to better understand the contribution of gel microstructure to rheology, we investigate the microstructure of the gel of attractive rods during and after shear cessation using combined rheology and confocal microscopy. Due to experimental challenges, similar measurements could not be satisfactorily carried out for gels of attractive spheres and will be pursued in future. 

\subsection{\label{level3.5}Structure and rheology}

We captured three dimensional image stacks (50 $\times$ 50 $\times$ 50 $\mu$m$^3$) of the gel microstructure after shear cessation. Images were captured after allowing the same aging time with DTS measurements (t$_w$ = 2000 seconds). This measurement was restricted to the gel with lowest rod volume fraction ($\phi$ = 0.04) studied here due to the large quantity of fluorescent particle suspension required for the size of the cone-plate geometry used in the study. Similar to the observation in our previous study,\cite{das2020shear} there was no change in gel microstructure during aging which is in agreement with the phenomenon of contact driven aging in colloidal gels formed by screening Coulomb interactions.\cite{pantina2005elasticity,pantina2006colloidal,bonacci2020contact}
\begin{figure}[hbt!]
	\centering
	\includegraphics[width=1\linewidth]{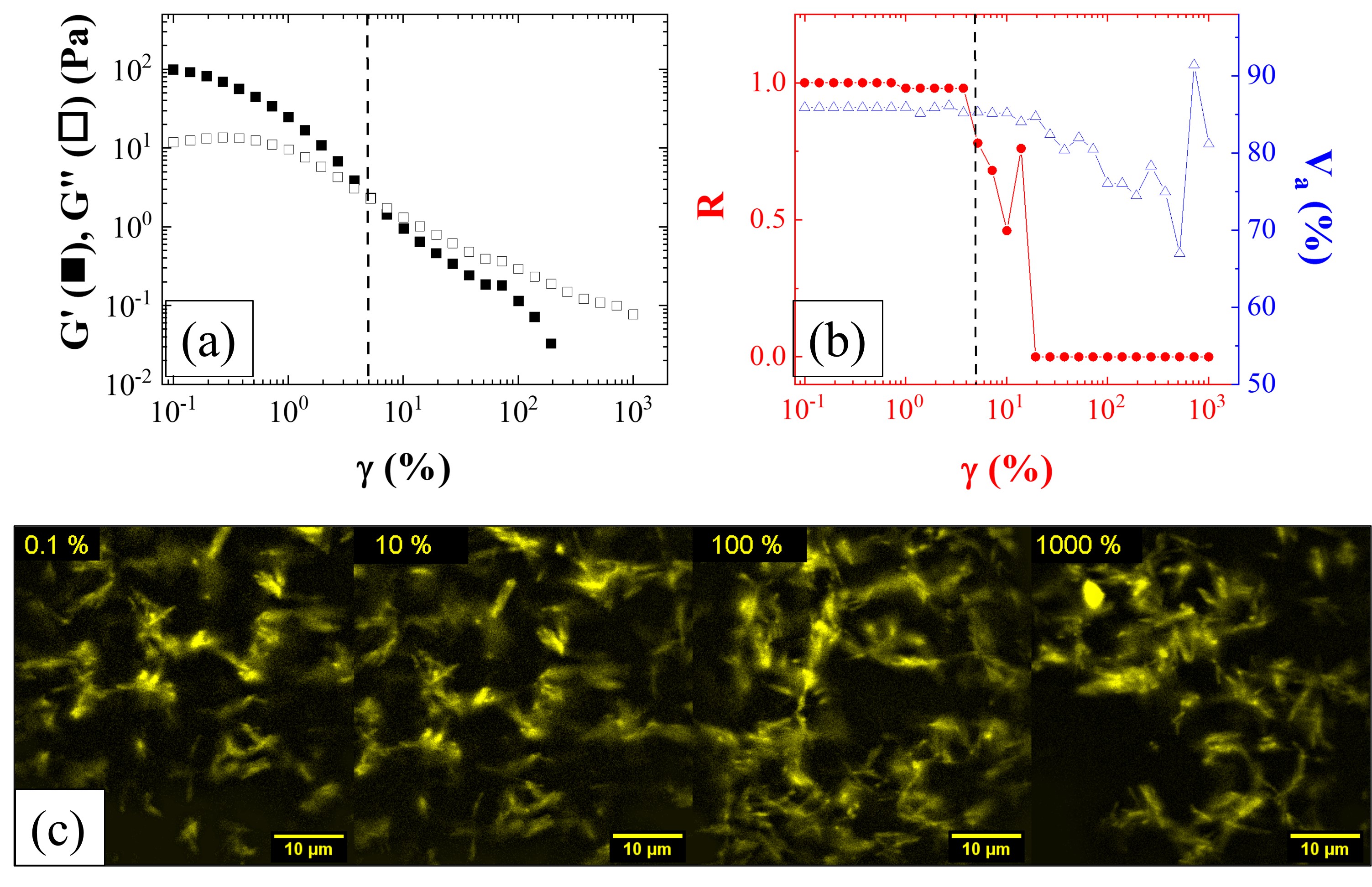}
	\caption{(a) DSS measurement ($\omega$ = 1 rad s$^{-1}$) of silica rod gel ($\phi$ = 0.04) and (b) the corresponding image correlation coefficient (R) calculated from correlating successive 2D confocal micrographs with that of gel at quiescent state and void area (V$_a$) calculated from the same confocal micrographs. (c) Confocal micrographs captured at a height of 25 $\mu$m from the bottom wall during DSS measurement.}
	\label{fig:fig12}
\end{figure}
\par We first captured the microstructure of gel of attractive rods at a height of 25 $\mu$m from the bottom wall while they underwent oscillatory flow during DSS measurement. Only two dimensional images were captured during this measurement for faster scan rate. We used an image cross correlation technique which is similar to the widely used two dimensional digital image correlation (2D DIC) method in experimental mechanics for capturing the microscopic structural changes as a material undergoes deformation.\cite{sutton2009image} In this method the grey intensities from the digital image of a surface before and after deformation are compared.\cite{pan2009two} In our case, we cross correlate the successive 2D confocal images captured during DSS measurement with the 2D confocal image of the gel microstructure at quiescent state using a plugin (\textit{ImageCorrelationJ$\_$1o})\cite{chinga2007quantification} in ImageJ software. The plugin determines Pearson correlation coefficient R which provides linear degree of correlation between two series of images. More details regarding this measurement can be found in the Appendix. For fully correlated images, the value of R = 1 and for fully decorrelated images the value of R = 0.
\par We also determined the percentage void area in binarized 2D confocal images. Binarizing is done after performing a thresholding of the intensity distribution in a 2D image with a cut-off grey scale value which is chosen empirically to best represent the image details after thresholding. After binarizing the images, value of pixels with presence of rods are counted 1 and with the voids as 0 and the total area of voids in a 2D image was calculated. Fig. \ref{fig:fig12} shows (a) the DSS measurement of gel of attractive rods and (b) the corresponding image correlation coefficient R and percentage void area (denoted as V$_a$) in the gel microstructure. At the bottom (c) we can see the confocal micrographs captured at different strain amplitudes. Even though this analysis does not provide all microstructural details during yielding, it does offer some information about local re-arrangements within the gel network in a specific area.
\par Here we observe that G' starts to drop at $\gamma$ as low as 0.3 \% while there is no change in values of R (Fig. \ref{fig:fig12} (b)). The value of R drops close to $\gamma \approx$ 4 \% which is the cross-over strain, $\gamma_{c}$ (G' = G''). The absence of structural changes at $\gamma < \gamma_{c}$ while G' exhibits continuous drop indicates that there exists some some local interparticle breaking or rearrangement that precedes microstructural transition and is not detectable by R or V$_a$ . The value of R drops to zero already around $\gamma$ = 10 \% indicating complete decorrelation of the images of the gel microstructure at this $\gamma$ with that at quiescent state. Another important observation made here is that at intermediate strain amplitude $20\% < \gamma < 200\%$, the gel network is observed to bend, deform and compactify rather than rupture which results in a sharp drop in V$_a$ as a large compact cluster occupies the field of view. On further increasing $\gamma$ above 300\% we observe rupturing of the gel network and movement of large rod clusters in and out of the image field of view. In addition, these observations make it clear that in semi-dilute suspensions of rods having strong short-range attractions, large strain amplitudes do not result in flow-alignment of individual rods.
\par
\begin{figure}[hbt!]
	\centering
	\includegraphics[width=0.7\linewidth]{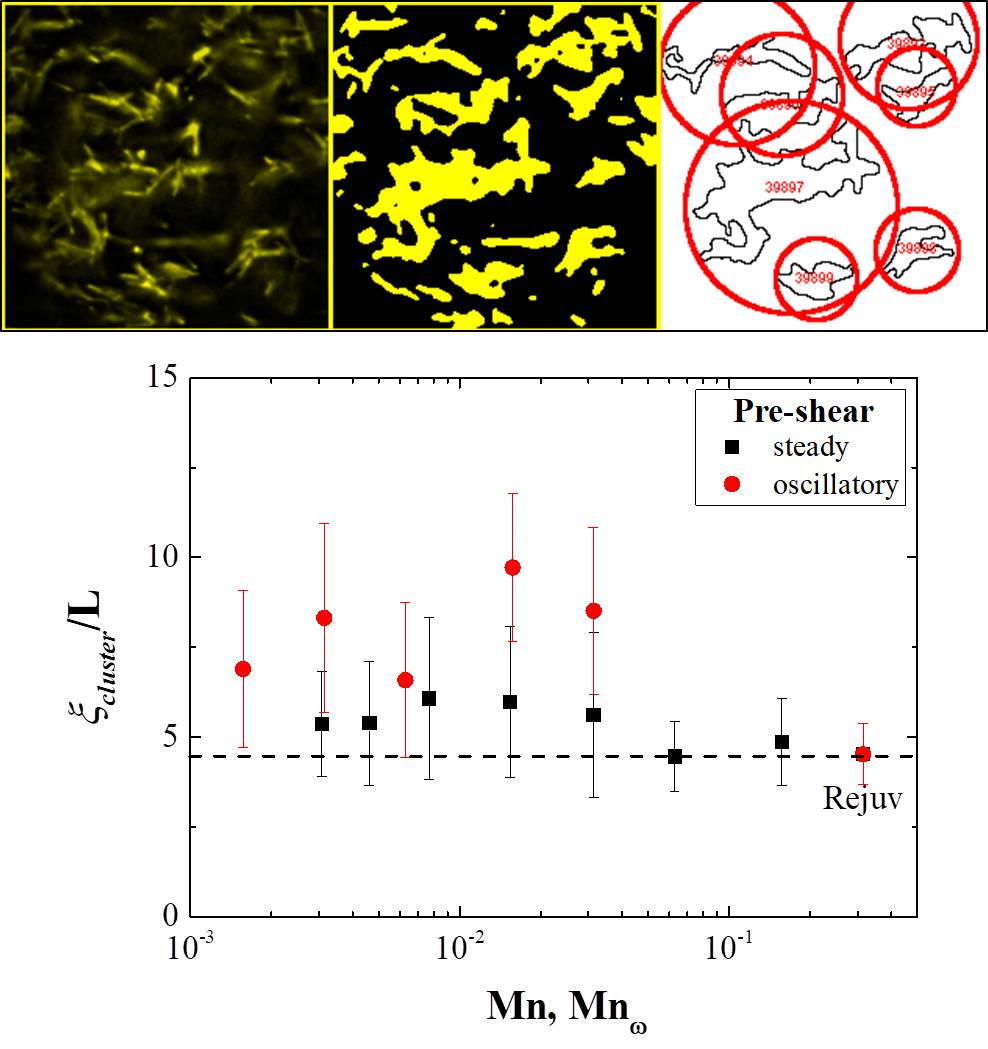}
	\caption{(Top) Steps in determination of particle cluster size in gels of attractive rods, red circles indicate the Feret's diameter, (bottom) particle cluster size $\xi_{cluster}$ of gel microstructure normalized by rod length L vs. Mn from steady and oscillatory pre-shear.}
	\label{fig:fig13}
\end{figure}

 Next we look at the three-dimensional microstructure of the gel formed by attractive rods after cessation of different pre-shear protocols (t$_w$ = 2000 seconds). Here the gel microstructure forms a space spanning isotropic network of rods immediately after cessation of the highest shear rate used for rejuvenation (Mn = 0.3). As was explained earlier, we do not observe any changes in particle microstructure during the aging process unlike the case of colloid polymer gels.\cite{dinsmore2001three} Two dimensional confocal micrographs reveal clear  microstructural changes associated with pre-shear. Despite the type of flow associated with oscillatory or steady pre-shear protocols, overall the gel maintains isotropic nature of the rods within the particle cluster without any indication of local nematic ordering. Moreover, when comparing steady and oscillatory pre-shear at intermediate Mn, we observe a compact rod cluster formed under oscillatory pre-shear. By comparing two dimensional and three dimensional confocal micrographs we observe that these compact clusters are large enough to span across the sample volume observed. This is an indication of cluster densification under oscillatory shear.\cite{mohraz2005orientation,laurati2011nonlinear,koumakis2011two} We measured the cluster size $\xi_{cluster}$ throughout the three dimensional image stack of the gel microstructure. $\xi_{cluster}$ was determined by first binarizing the images and then determining the Feret's diameter (distance between furthest points on an irregular shape) of a cluster. Only clusters with a diameter greater than a double rod length were considered. The analysis was performed on each 2D confocal image in a 3D image stack to determine the average cluster size in the gel microstructure. The cluster size normalized by rod length was plotted against the pre-shear rate or Mn (Fig. \ref{fig:fig13}). We see that the average cluster size after oscillatory pre-shear is larger compared to steady pre-shear indicating larger structural heterogeneity. Our results are in agreement with previously reported study by Moghimi \textit{et al.}\cite{moghimi2017colloidal} where oscillatory pre-shear introduces larger heterogeneity in colloidal gel microstructure indicative of more efficient shear protocol for tuning their viscoelasticity.
\par We should note that  a similar structure-rheology relationship for gel formed by attractive spheres ($\phi$ = 0.25) is not that sucessful due to dense nature of the gel and poor refractive index matching between silica particles and 11 M CsCl, with the gel microstructure not being clearly visible beyond 5 $\mu$m from the bottom wall. 

\section{\label{level4}Discussion}
We now summarize and discuss the effects of different pre-shear protocols on the gels of rods or spheres (with similar attraction) at different volume fractions. Therefore as demonstrated above, applying different pre-shear leads the metastable system to distinct long-lived minima of the energy landscape. Practically this enables us to create samples with the same composition of ingredients but with different structural and mechanical properties.
\par \textit{\textbf{Bulk modulus and gel-wall interactions}}: Flow curves of both gels exhibit a finite yield-stress plateau (Fig. \ref{fig:fig2}). However, in the case of gels of attractive rods, we observe stronger wall-slip (shown by drop in stress at low shear rates compared to gels of attractive spheres). Wall-slip in colloidal gels is associated with reduced number of contact points between particles and the wall as a result of growing heterogeneity and particle cluster size at low shear rates. This indicates that structural heterogeneity at low shear rates is higher in gels of attractive rods compared to attractive spheres. Linear rheological measurements reveal that at quiescent state, gels of attractive spheres exhibit a higher power law slope for G' with increasing particle volume fraction compared to that of rods. This could be due to reduced number of particles and contact points within a homogeneous gel network of low volume fraction rods compared to high volume fraction spheres. Reduced number of particles and contact points means less number of stress bearing elements within the gel network leading to a weaker gel. Moreover, we expect some contribution from particle surface roughness.
\par Aging after rejuvenation shows both types of gel exhibiting solid-like response almost immediately  after shear cessation. One of the reasons for this is that dense colloidal gels formed by particles with van der Waals attraction are able to percolate in a short time span ($<$ 6 seconds). In the case of rods, they have larger excluded volume compared to spheres and are able to percolate at comparatively lower volume fractions. As particles diffuse freely and find a neighbour, they form quite strong bonds as the strength of attraction U $\approx$ -45 \textit{k$_B$T} after which the particles are permanently arrested. The second reason for the fast gelation after rejuvenation is that the highest Mn achieved is less than 1 (Mn = 0.3). Under such condition, it is not possible to break all the bonds in order to disperse individual particles. Hence, after shear cessation we have percolation of smaller particle clusters (both in spheres and rods) which is much faster at high particle volume fraction than percolation of individual particles. Moreover, the gel network stiffens over time due to contact-controlled aging. \cite{bonacci2020contact} Such aging effects are expected to affect not only bulk properties but the gel - wall interactions, although this has not been explored here.
\par\textbf{\textit{Particle anisotropy effects in the microscopic yielding}}: The broadness in the elastic stress peak in DSS measurements of pre-sheared rod gels, shows that the yielding process is gradual with multiple bond-breaking events occurring simultaneously before the network ruptures. As the bonds between isotropic rod networks break, some rods are free to rotate around their contact points leading to flexibility in the gel network and formation of new bonds with decreasing probability before rupture at larger strain amplitude.\cite{das2020shear} It has been shown that a DLCA cluster of hard-sphere gel exhibits similar flexibility due to unwinding of the cluster before rupture.\cite{mohraz2005orientation,chan2012two} A similar effect takes place in gels of attractive rods at intermediate strain amplitudes during oscillatory pre-shear which can be seen as a repeating cycle of yielding and reformation processes. This is also evidenced by direct imaging as observed in the confocal microscopy videos during LAOS. At these strain amplitudes, the rod network is partially ruptured and does not have enough time to reform. This allows the rod clusters to compactify over time (as observed in confocal images) by continuously undergoing intra-cycle structural transitions. At lower strain amplitudes, this effect is much smaller and does not significantly increase the structural heterogeneity while at very high strain amplitudes the gel network is broken into small rod clusters which are uniformly distributed throughout the sample volume.
\par Our study shows that the yielding of gels of particles (spheres or rods) with van der Waals attraction involve inter-cluster bond breaking at $\gamma_0 \approx \gamma_{c}$, however, does not involve intra-cluster bond breaking which is associated with two-step yielding as observed in colloid-polymer depletion gels.\cite{laurati2011nonlinear,koumakis2011two,chan2012two} Furthermore, in the present study we have not reached Mn $\gg$ 1, necessary for full intra-cluster breaking.
\par\textbf{\textit{Training and tuning of the gel - Effects of particle anisotropy}}: The pre-shear study reveals that after steady pre-shear at intermediate and low Mn (see Fig.\ref{fig:fig9} (a)), gels of attractive spheres exhibit strengthening before exhibiting sedimentation at very low Mn ($<$ 10$^{-3}$) whereas this effect was almost non-existent in the case of gels of attractive rods. This indicates that structural heterogeneity in gels of attractive spheres at such Mn leads to formation of a stronger gel (except for $\phi$ = 0.25), a process known as over-aging where shear assists the colloidal system to age faster. However, at very low Mn, formation of very large particle clusters can lead to structural collapse due to density mis-match leading to sedimentation. We believe as reported by Moghimi \textit{et al.},\cite{moghimi2017colloidal} that structural heterogeneity in gels of attractive spheres would be even higher under LAOS pre-shear as is evident by sedimentation at $\gamma_0 < \gamma_{c}$ which can arise as a result of the creation of large particle clusters. Maximum strength of gels of attractive spheres after LAOS pre-shear is observed at $\gamma_{0}$ slightly above $\gamma_{c}$. This indicates that by shearing the gel of attractive spheres close to $\gamma_{c}$, it is possible to form a stronger gel than at quiescent state. Such a process would allow particles to weakly reorganize themselves into configurations to minimize their free energy as is observed during aging.\cite{joshi2015model} Furthermore, the elastic stress peak (or yield stress) shifts to lower strain amplitudes with reduction in pre-shear rate or Mn (see Fig. \ref{fig:fig11}). This is indicative of stiffening of the gel network. The pre-shear thus leads here both to an increase of the elastic modulus and a decrease of the yield strain, thus creating a stronger yet more brittle structure.\cite{colombo2014stress} The pre-shear process can thus be viewed as ``training'' of the colloidal gel where we employ a shear-assisted aging by performing finite cyclic deformation for the gel to reach its minimum free energy state. Oscillatory shear close to $\gamma_{c}$ is more effective compared to low steady shear rate in tuning the gel elasticity as the former does not involve large structural deformation. It is also possible that in the absence of sedimentation, an even stronger gel might be obtained by shearing at $\gamma_0 \approx \gamma_{y}$ instead of $\gamma_{c}$, a regime we plan to study in the future with more carefully density matched systems.
\par Gels of attractive rods are less prone to strong sedimentation effects. Since $\phi_{rcp}$ of isotropic rods (= 0.54) is much lower than spheres (= 0.64),\cite{philipse1996random} cluster density of gels of isotropic attractive rods will be much lower compared to spheres leading to increased stability. Furthermore, due to their larger excluded volume, rod clusters percolate despite the presence of large voids arising from structural heterogeneity at low Mn preventing sedimentation. Rheo-confocal results reveal that structural reorganization in rod gel takes place at $\gamma_0 \approx \gamma_{c}$ beyond which cluster densification occurs. These densified clusters cause wall-slip at larger $\gamma_0$ leading to drop in stress. A similar drop is also observed for gels of attractive spheres. 
\section{\label{level5}Conclusions}
In summary, we have investigated the effect of different pre-shear protocols (steady and oscillatory) on the mechanical response and structure of colloidal gels made up of attractive silica spheres or rods suspended in 11 M CsCl. Gels of attractive spheres exhibit strengthening at intermediate pre-shear rates (or Mn) and weakening and collapse at lower pre-shear rates for the higher volume fraction. However, at lower volume fraction gels they  also show a weakening at intermediate rates similar to the depletion system,\cite{moghimi2017colloidal} before they strengthen again at lower strain amplitudes. On the other hand, rod gels exhibit weakening under both types of pre-shear with the exception of mild strengthening at intermediate shear rates after oscillatory pre-shear. The strengthening effects in gels of attractive spheres are attributed to shear -assisted structural rearrangements within the gel network that leads to over-aging or evolution of the structure to attain minimum free energy. Rheo-confocal studies reveal that heterogeneity in microstructure of both type of gels changes with applied pre-shear rates and the isotropic nature of the rod gel is retained under all pre-shear conditions. At intermediate Mn under oscillatory pre-shear, gels of attractive rods form more compact particle clusters. Consequently, gels of attractive rods are weaker after oscillatory pre-shear compared to steady pre-shear. The yielding process of gels of attractive spheres is sharper whereas for attractive rods, it is more gradual. We believe that in gels of attractive spheres, this is due to inter-cluster bond breaking, whereas in attractive rods it is due to multiple bond breaking events allowing individual rods to rotate imparting flexibility to gel network before rupture.
\par A clear tunability of gels of attractive spheres and rods is achieved through different mechanisms. Further studies with the help of simulations are required to obtain quantities such as bond numbers, local rod orientation, void distribution, etc. to explain better the factors behind the difference in their tunability. Moreover, further rheo-confocal studies with better refractive and density matched systems such as PMMA depletion gels may provide better quantitative results in terms of structure-rheology relationship.
		
\begin{acknowledgments}
	We acknowledge the support from Marie-Sklodowska Curie Innovative Training Network ``DiStruc" (Grant agreement no. 641839) as well as innovation program ``EUSMI" (Grant agreement no. 731019). We would also like to acknowledge L. Chambon, E. Vasilaki and M. Vamvakaki, Univ. Crete for providing the particles.
\end{acknowledgments}

\appendix
\section{Image Cross-correlation}
\label{Appendix}
Image cross-correlation was performed on greyscale images acquired during rheo-confocal measurements. Successive confocal micrographs captured during yielding under LAOS were cross-correlated with the confocal micrograph captured for the rod gel microstructure at quiescent state. Pearson correlation coefficient is given by,\cite{pearson1895vii}
	\begin{equation}
		\label{eq:eq4}
		R = \dfrac{\sum^{m}_{i=1}\sum^{n}_{j=1} ((f(m_i n_j)-\stackrel{-}{f}))((g(m_i n_j)-\stackrel{-}{g}))}{(mn-1)\sigma_f \sigma_g}
	\end{equation}
	Here \textit{m} and \textit{n} are the number of sub-regions (size 3$\times$3 pixels in our case) in \textit{x} and \textit{y} directions respectively out of a larger 2D image (253$\times$253 pixels). \textit{f(m$_i$ n$_j$)} and \textit{g(m$_i$ n$_j$)} are the values of intensities of sub-regions at position \textit{ m$_i$} and \textit{n$_j$} in the first and second image respectively. $\stackrel{-}{f}$ and $\stackrel{-}{g}$ are the mean, $\sigma_f$ and $\sigma_g$ are the standard deviation of the intensities of first and second image respectively and are given by,
	\begin{equation}
		\label{eq:eq5}
		\stackrel{-}{f} = \left(\dfrac{1}{mn}\sum^{m}_{i=1}\sum^{n}_{j=1}f(m_i n_j)\right)
	\end{equation}
	\begin{equation}
		\label{eq:eq6}
		\sigma = \left(\dfrac{1}{mn-1}\sum^{m}_{i=1}\sum^{n}_{j=1}(f(m_i n_j)-\stackrel{-}{f})^2\right)^{(1/2)}
\end{equation}
For fully correlated images, the value of R = 1 and for fully decorrelated images the value of R = 0. Further information regarding a detailed analysis of this method is reported elsewhere.\cite{chinga2007quantification}
		
		\nocite{*}
		\bibliography{tuning_gel}
		\bibliographystyle{unsrt}

	\end{document}